\documentclass[12pt, a4paper]{article}
\usepackage{tikz}
\usetikzlibrary{matrix,shapes,arrows,positioning,chains}
\usepackage{graphicx}
\usepackage{float,bm,afterpage,amsmath,placeins,authblk,adjustbox,lscape,caption}
\usepackage[english]{babel}
\usepackage{setspace}
\usepackage{booktabs}
\usepackage{changes}
\usepackage[top=1.3in,left=1.1in,right=1.1in,bottom=1.3in]{geometry}
\usepackage{color}
\usepackage{hyperref}
\graphicspath{{figures/}}

\date{}

\begin{document}

	\title{Filtering Procedures for Sensor Data in Basketball}
	
	\author[1]{Rodolfo Metulini}

	\affil[1]{\small Department of Economics and Management, University of Brescia}

	\maketitle
	
	\textbf{Please cite as:} \textit{Metulini, R. (2017), Filtering Procedures for Sensor Data in Basketball, Statistics \& Applications. Vol. 2.}

\begin{abstract}
Big Data Analytics help team sports' managers in their decisions by processing a number of different kind of data. With the advent of Information Technologies, collecting, processing and storing big amounts of sport data in different form became possible. A problem that often arises when using sport data regards the need for automatic data cleaning procedures. In this paper we develop a data cleaning procedure for basketball which is based on players' trajectories. Starting from a data matrix that tracks the movements of the players on the court at different moments in the game, we propose an algorithm to automatically drop inactive moments making use of available sensor data. The algorithm also divides the game into sorted actions and labels them as offensive or defensive. The  algorithm's parameters are validated using proper robustness checks.
\end{abstract}


\section{Introduction}

Professional team sports' managers, more and more in recent years, are becoming aware of the potential of Data Analytics in order to better manage their team. In team sports, continuous interactions among three agents - coaches, single players and the whole team - produce an high level of complexity. This complexity has been studied, among others, in the new domain of ecological dynamics (\cite{travassos2013performance,Araujo2016team}). Nowadays, Information Technologies (IT) make large amounts of real-time information on teams and players available. Most of the results from the interaction among these three agents could be captured by two kind of data: (i) play-by-play data (also called $event-log$), which report a sequence of relevant events that occur during a match, related to either the team or the single player, such as shots or fouls; (ii) the positioning, the velocity and the acceleration of players or the ball, also called sensor data, which is captured through Global Positioning System (GPS) techniques. There is high potential in jointly using these two kind of data to cope with the intrinsic complexity in team sports and with the aim of producing advanced statistics for team managers. Several aspects are already taken into account in the scientific literature, \cite{gudmundsson2017spatio} being a nice and quite complete review. For example,\cite{carpita2013,carpita2015} used data mining techniques in order to identify the drivers that mostly affect the probability to win a football match. Social network analysis has also been used to capture the interactions between players (\cite{wasserman1994social}); \cite{passos2011networks} used centrality measures with the aim of identifying central players in water polo. A necessary condition to produce statistics is to correctly understand data, by collecting, storing and processing them in a proper way. A review on this regard has been made by \cite{passos2011networks}. 

This paper is about basketball data processing. Basketball is a sport  played by two teams of five players each on a rectangular court. The objective is to shoot a ball through a hoop 46 centimetres in diameter and mounted at a height of 3.05 meters to backboards at each end of the court. According to International Basketball Federation (FIBA) rules, the match lasts 40 minutes, divided into four periods of 10 minutes each. There is a 2-minute break after the first quarter and after the third quarter of the match. After the first half, there is an half-time break. 

The manuscript focuses on the processing of players' movements data.  In particular, the aim of this manuscript is to i) automatically drop all the inactive moments from a data matrix that tracks the movements of the players on the court at different moments in the game, ii) automatically split the game into sorted actions, labelling them as offensive or defensive. To do that we make use of available sensor data tracked during a game. This work is similar to that by \cite{wu2017modeling}, as they provide a procedure to process ball's and players' trajectories. However, our procedure differs, since it works even if data on ball's movement is missing. We place this piece of research within the domain of Human Activity Recognition (HAR). HAR aims to recognize the actions of an agent from a series of observations on the agents' actions and the environmental conditions, \cite{trabelsi2013approach} being a representative article on such a topic. In this work, the agents are considered to be the players of a team as a whole moving inside the court, and the action to be recognized concerns whether the game is active or inactive. In this vein,\cite{huang2012calculate} apply data automation algorithm in sports, by using sensor data to categorize golf swing trajectories. \cite{jiang2004new} propose a game segmentation algorithm that suits in different sports. \cite{jordan2009optimizing} propose a model based on design of experiments and response surface methodology.

In this paper we propose and discuss a multiple-stage algorithm which aims to drop inactive moments of a basketball data matrix tracking players' movements. We apply this algorithm to different real case studies (CS) in order to calibrate the algorithm's parameters by means of a data-driven approach. We then provide some descriptive statistics related to the CS for a validation check of the algorithm and of the robustness of the parameters.

In Section 2 we present and explain the algorithm. In Section 3 we validate the algorithm using real data. Section 4 concludes the paper and suggests further analysis.

\section{The algorithm}

Here we describe the algorithm aimed to automatically reduce a basketball data matrix with data on players' positions at different times to just the moments when the game is active, and to consistently split the game into sorted and labelled (offensive or defensive) actions, using tracked sensor data only. The algorithm is suitable in cases where i) information on the movement of players on the court has been captured with the use of appropriate GPS devices, for example, the accelerometer, a device that measures proper acceleration (\cite{tinder2006flight}); and ii) nobody, during the game, is in charge to take note of relevant informations of the game, such as active moments, offensive or defensive moments, and so on. Just to be clear, the only information available is that from GPS devices. 

As accelerometers track information of players' movement along the full game, data consists of a total of around 90-100 minutes, despite only 40 of these are actually related to moments of active play, and therefore relevant to the aim of the analysis of players' movement. For this reason, the objective of the algorithm is to reduce game data to the moments when the game is active (40 minutes). A parallel objective is to split the game into actions, in such a way the first (in chronological order) action is identified with a 1, the second one with a 2, and so on. In doing that, we obtain a reduced data matrix having a correct number of actions (each with a correct length duration).

\vline

The algorithm applies to a data matrix \textbf{X} where each row corresponds to a sensor record, described by the variables time (in milliseconds, $ms$). \textbf{X} should be sorted from the smallest $ms$ to the largest $ms$. Each row contains information on several variables related to the positioning ($pos$) and the velocity ($vel$) of each single player of one team ($p_1, p_2, ..., p_k$), along the court length ($x$), and the court width ($y$). Table \ref{tab:1} reports a sample of \textbf{X}. It is important to clarify that, with the accelerometer, data are detected with a non-constant frequency; in addition, data of different players are recorded at different time instants. The dataset should contain any detected time instant. To do that, is necessary to attribute the last datum available to players not detected in that time instant. This procedure will be described in detail in section 3.

\begin{table}[h!]	
	\small	
\caption{Sample of a subset of the data matrix \textbf{X}}
	\label{tab:1}
\begin{tabular}{ccccccccc}
ms & $pos_{p_1\_x}$ & $pos_{p_1\_y}$ & $vel_{p_1\_x}$ & $vel_{p_1\_y}$ & $pos_{p_2\_x}$ & $pos_{p_2\_y}$ & $vel_{p_2\_x}$ & $vel_{p_2\_y}$  \\
	\hline
	5564 & 4.28 & 7.40 & 1.26 & 1.26 & 15.25 & 8.98 & 0 & 0 \\
	5579 & 0.32 & 1.03 & 0.36 & 0.36 & 15.25 & 8.98 & 0 & 0 \\		
\end{tabular}
\end{table}

The algorithm consists of two main parts. In the first part it removes rows from \textbf{X} according to three different criteria. All these criteria are based on players' positioning and velocity. According to the first criterion the algorithm drops, from \textbf{X}, the instants in which the number of players inside the court is different from five. With this criterion we ideally remove all the moments related to pre-match and post-match periods, half-time and quarter-time intervals, time-outs and so on. The second criterion aims to drop, from \textbf{X}, the instants in which a player is shooting a free throw, by looking to his positioning in the court. The third criterion aims to remove those moments where all the five players in the court report a velocity lower than $h_2$ $km/h$, for at least $h_3$ consecutive seconds, where $h_2$ and $h_3$ are subject to a parameter calibration. The second part of the algorithm assigns actions' sorting and labelling to the reduced data matrix, by looking to the average positioning of the five players on the court.

\vline

We now describe the steps of the algorithm in detail, as also schematically summarized in Figure \ref{fig:1}.
\tikzset{
	decision/.style={
		diamond,
		draw,
		text width=4em,
		text badly centered,
		inner sep=0pt
	},
	block/.style={
		rectangle,
		draw,
		text width=10em,
		text centered,
		rounded corners
	},
	cloud/.style={
		draw,
		ellipse,
		text width=22em,
		minimum height=2em
	},
	descr/.style={
		fill=white,
		inner sep=2.5pt
	},
	connector/.style={
		-latex,
		font=\scriptsize
	},
	rectangle connector/.style={
		connector,
		to path={(\tikztostart) -- ++(#1,0pt) \tikztonodes |- (\tikztotarget) },
		pos=0.5
	},
	rectangle connector/.default=-2cm,
	straight connector/.style={
		connector,
		to path=--(\tikztotarget) \tikztonodes
	}
}

\begin{figure}[h!]
	\centering
	\begin{tikzpicture}
	\matrix (m)[matrix of nodes, column  sep=2cm,row  sep=8mm, align=center, nodes={rectangle,draw, anchor=center} ]{
		|[block]| {Full data matrix \textit{X} (nrow = T)};               &  \\
		|[cloud]| {\textbf{1-A}  Remove row if players on the court $\neq$ 5
		}               &                                            \\
		|[cloud]| {\textbf{1-B}  Remove row if a player is on the free throw circle
		}          &                                             \\
		|[cloud]| {\textbf{1-C}  Remove row if players velocity $<h_2$ for $h_3$ consecutive seconds}
		&                                             \\
		|[block]| {Reduced data matrix (nrow = T' $\le$ T) 
		}         &                                             \\
		|[cloud]| {\textbf{2-A} Assign offense or defense labels}    &                                             \\
		|[cloud]| {\textbf{2-B} Assign actions' sorting}         &                                             \\
		|[block]| {Reduced data matrix with actions' labelling and sorting}    &       \\                                       
	};

	\path [>=latex,->] (m-1-1) edge (m-2-1);
	\path [>=latex,->] (m-2-1) edge (m-3-1);
	\path [>=latex,->] (m-3-1) edge (m-4-1);
	\path [>=latex,->] (m-4-1) edge (m-5-1);
	\path [>=latex,->] (m-5-1) edge (m-6-1);
	\path [>=latex,->] (m-6-1) edge (m-7-1);
	\path [>=latex,->] (m-7-1) edge (m-8-1);
	\end{tikzpicture}
	\caption{Flow chart representing the steps of the algorithm.} 
	\label{fig:1}
\end{figure}
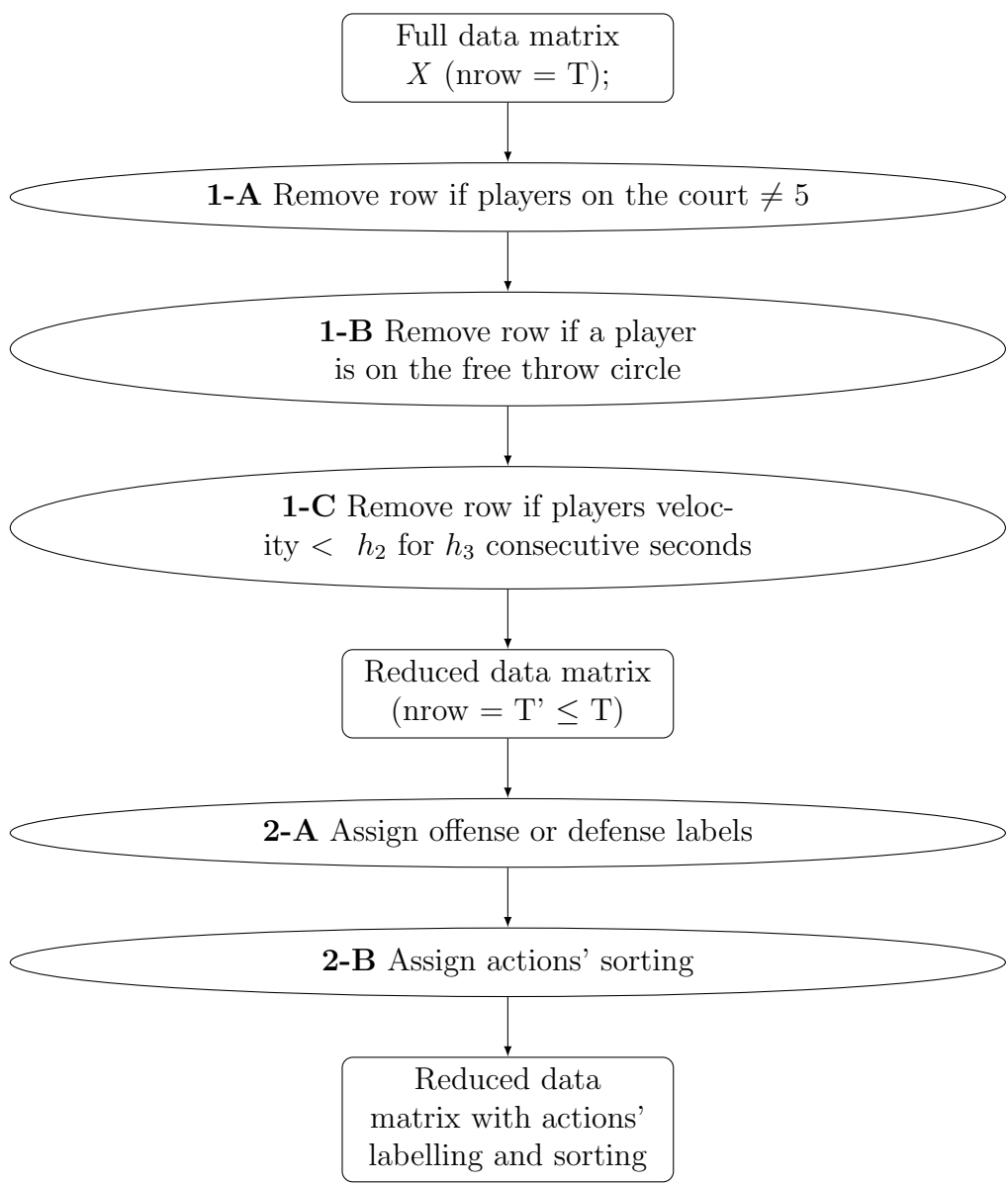

\begin{itemize}
	\item In the step 1-A th algorithm drops the rows where the number of players in the court is different from five. To do that, the algorithm creates a new variable for each of the k players. For player 1 ($p_1$), this variable assumes value 1 when the player's coordinates ($pos_{p_1\_x}$ and $pos_{p_1\_y}$) are inside the court, 0 otherwise. The algorithm, after having computed this variable for all the k players, creates a new variable ($count$), which is the row sum of the k 0/1 variables. If $count = 5$, it means that, in that specific $ms$, the number of players in the court is exactly five. The algorithm removes the rows where $count \neq 5$ (See Figure \ref{fig:2}).

	\begin{figure}[h!]
		\centerline{ \includegraphics[width=1\linewidth]{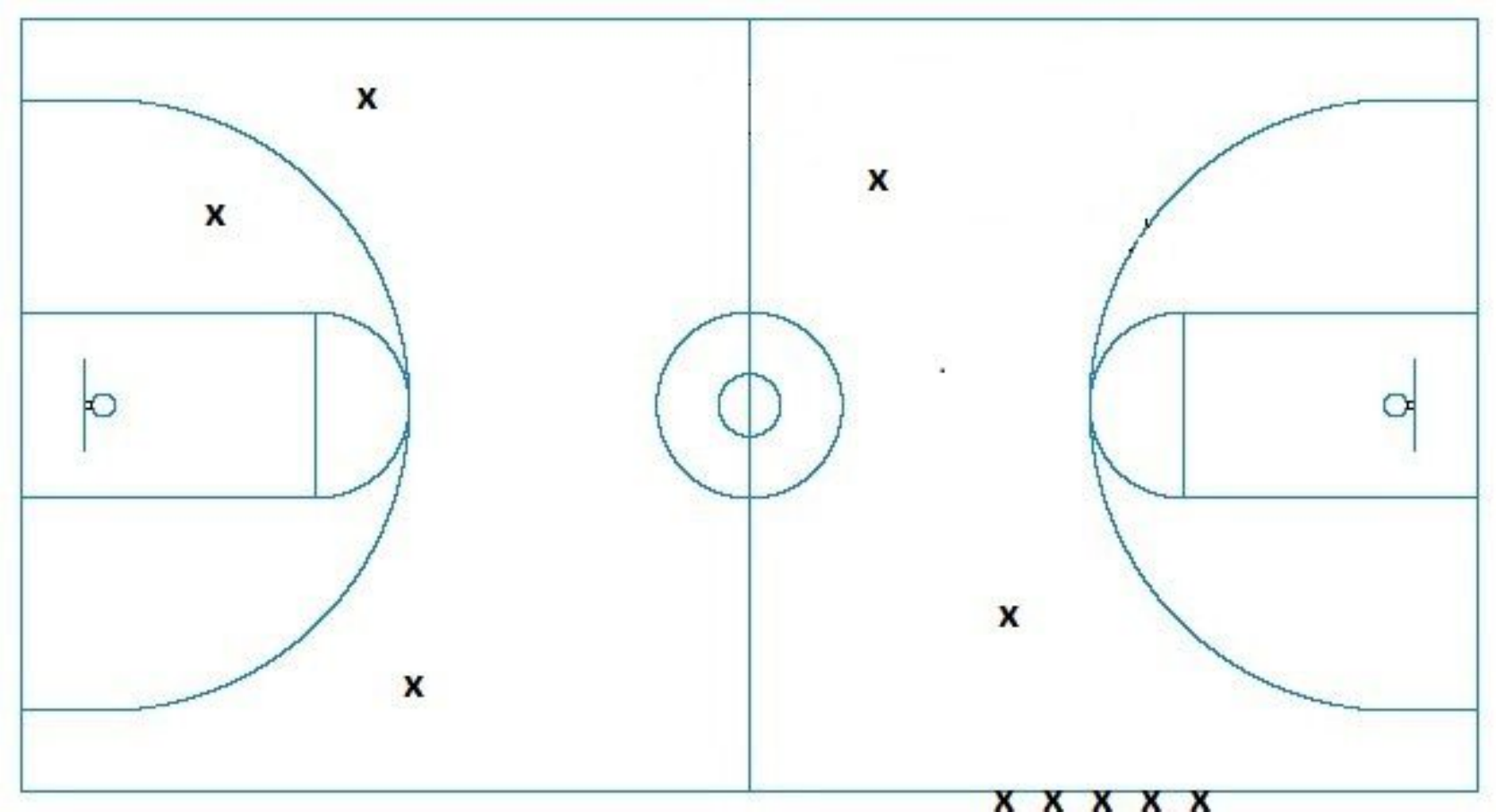}}
		\caption{\textit{step 1-A of the algorithm. The figure represents a moment in which exactly five players are inside the court.}}
		\label{fig:2}       
	\end{figure}
	
	\item In the step 1-B the algorithm drops the rows that correspond to the moments in which a player is shooting a free throw. To do that the algorithm generates a new variable ($ft$) which assumes value 1 when at least one player's coordinates lie inside the free throw circle. The algorithm assumes that a player is shooting a free throw when he remains inside the circle for at least $h_1$ = 10 consecutive seconds\footnote{We set parameter $h_1$ as a constant. However, a tuning could be apply on this parameter.}. 
	When $ft$ consecutively reports value 1 for at least 10 seconds, the algorithm drops all the corresponding rows from the data matrix (Figure \ref{fig:3}).

		\begin{figure}[h!]
	\includegraphics[width=0.5\linewidth]{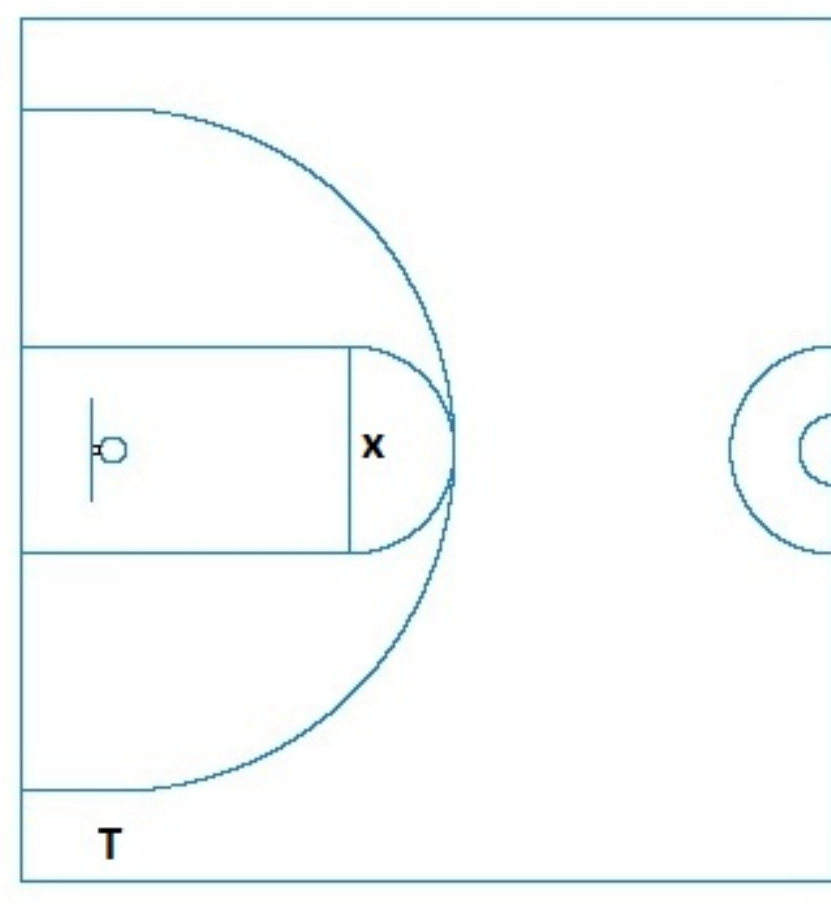}
	\includegraphics[width=0.5\linewidth]{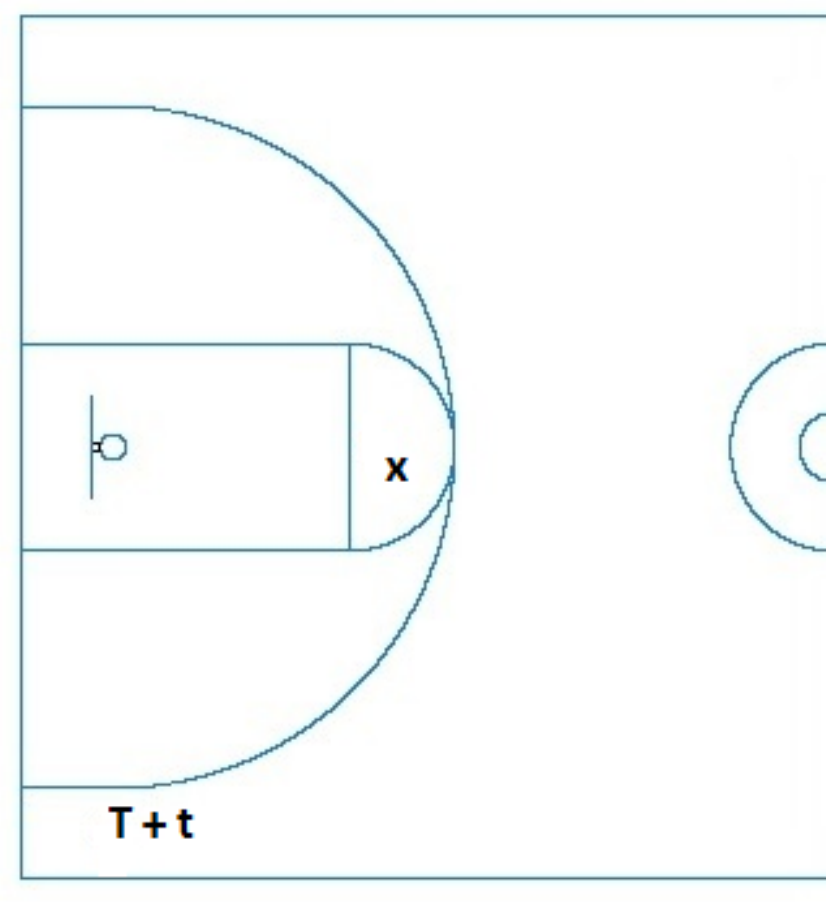}
	\caption{\textit{Step 1-B of the algorithm: the figure represents the criterion used to drop moments in which a player lies inside the free throw circle.}}
	\label{fig:3}       
\end{figure}

	\item In the step 1-C the algorithm drops the rows corresponding to moments where all the five players in the court are not running, for a certain number of consecutive seconds. This further step is necessary because steps 1-A and 1-B do not completely filter out inactive moments. For example, the moments in which the referee whistles for a foul, the moments when the ball comes back into play, or the moments of players' change are not detected in the first two steps. Moreover, since trajectories are available for one team only, step 1-B drops moments in which a player from that team is shooting a free throw, while step 1-C should also drops moments when a free throw has been attempted by a player of the opponent team. To do that, the algorithm first computes the velocity of each player, being $vel_{p_1} = \sqrt{vel_{p_1\_x}^2 + vel_{p_1\_y}^2}$ the velocity of player 1. There are moments where all the five players' velocity is less than $h_2$ $km/h$ for at least $h_3$ seconds.  Those are the moments to be dropped. Thinking to a real game, we assume a feasible range for the parameter $h_2$ to be larger than 8, whereas this measure is expressed in $km/h$; in fact, a walking player does not exceed 8 $km/h$. Passing to $h_3$, a feasible range for this parameter is larger than 1, as there could be active moments (lower than 1 second) where nobody is running.
\end{itemize}

\begin{figure}[h!]
	\includegraphics[width=1\linewidth]{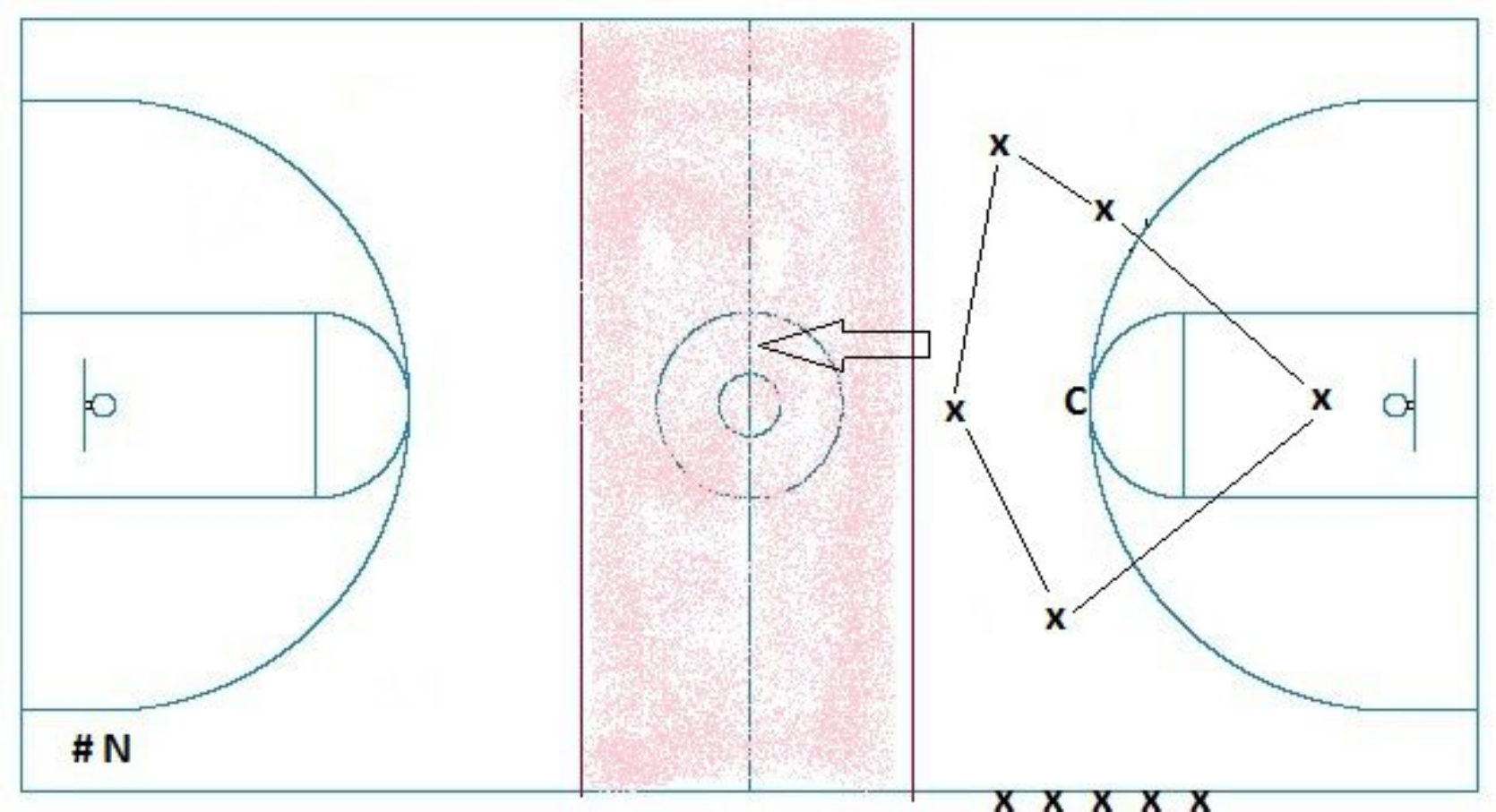}
	\includegraphics[width=1\linewidth]{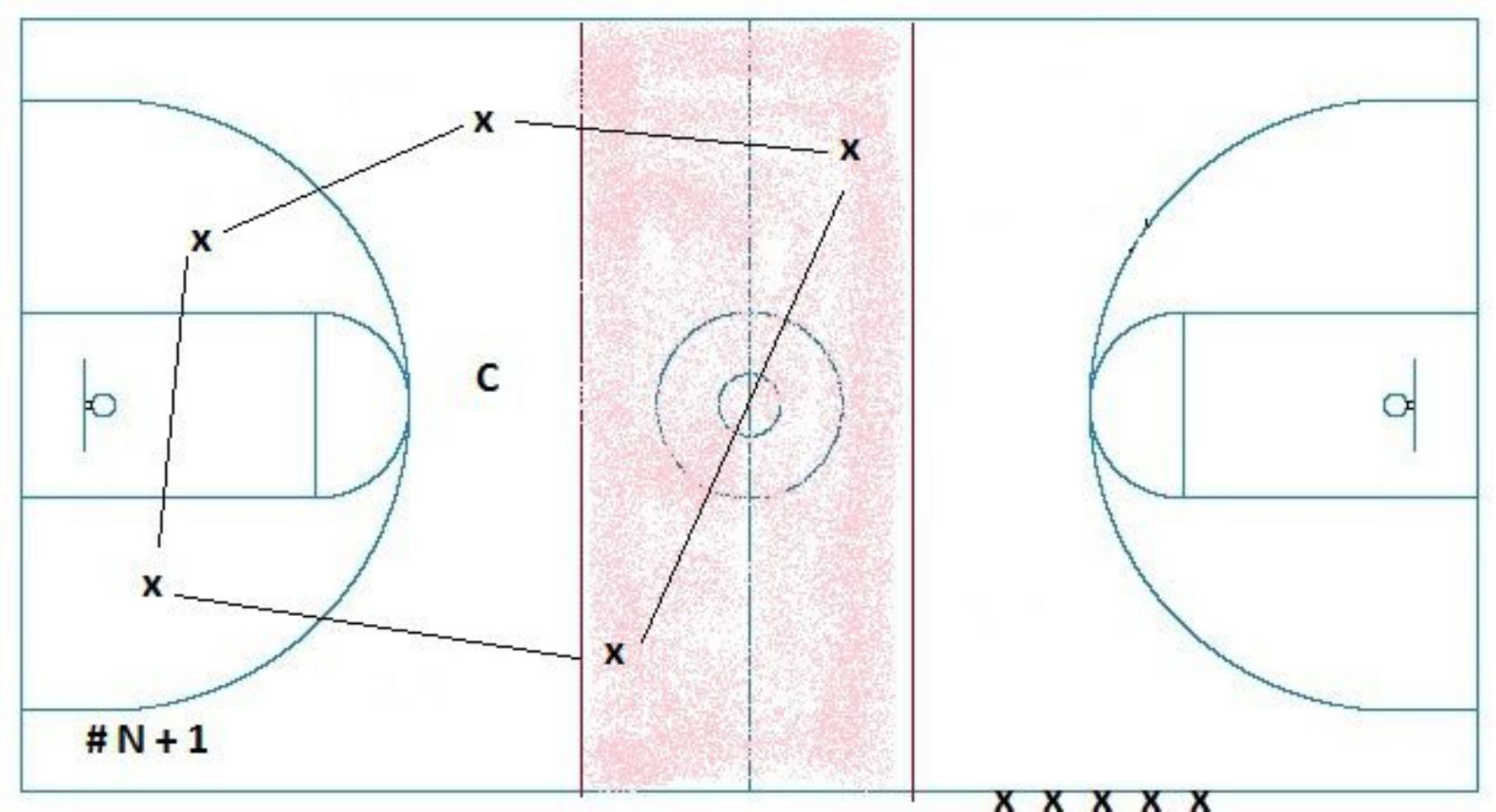}
	\caption{\textit{Step 2-B of the algorithm: figures represent the criterion used to assign moments to consecutive actions.}}
	\label{fig:5}       
\end{figure}

After these three steps, the algorithm generates a reduced version of the data matrix, which should include information for about 40 minutes of game. The reduced version is then processed throughout these two additional steps: 
	
\begin{itemize}
	\item The step 2-A aims at assigning a label to each row of the data matrix. The label regards whether the row belongs to a moment where the team is in offense or in defense. 
	In doing that, the algorithm generates a new variable on the reduced version of the matrix \textbf{X}, $avg\_pos$, that represents the average coordinate of the five players on the court, where $avg\_pos$ is a vector of two elements $[avg\_pos_x;avg\_pos_y]$, being $avg\_pos_x = \frac{\sum_{i=1}^{5}pos_{p_i\_x}}{5}$ and $avg\_pos_y = \frac{\sum_{i=1}^{5}pos_{p_i\_y}}{5}$. $avg\_pos$ could lies either on the offensive or on the defensive side of the court (see Figure \ref{fig:4}). The algorithm also labels transition, that corresponds to those moments having $avg\_pos_x$ in the interval [+4,-4] meters ($m$) from the half-court line. The step ends by generating a new variable that assumes either value $O$ (offense), $D$ (defense) or $Tr$ (transition)\footnote{The algorithm takes into consideration that after the half-time break the two teams change court side.}. 
	
	\begin{figure}[h!]
		\includegraphics[width=1\linewidth]{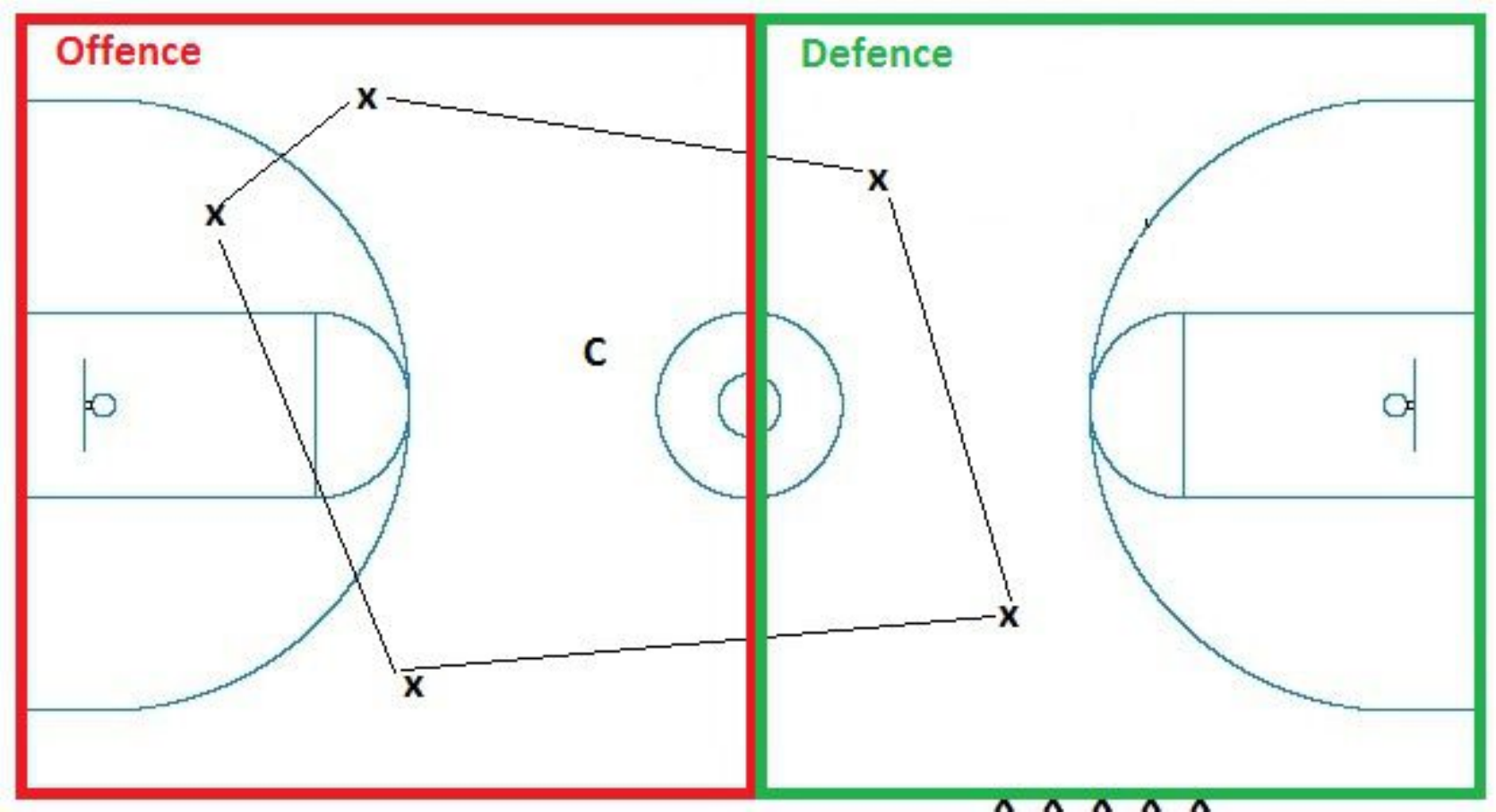}
		\caption{\textit{Step 2-A of the algorithm: the figure represents the criterion used to assign ``offense'' or ``defense'' labels.}}
		\label{fig:4}       
	\end{figure}
	
	\item In the step 2-B the algorithm attributes to each row its action number, so that the moments related to the first action report value 1, the moments of the second action report value 2, and so on. We adopt the following criterion: the algorithm creates a new variable ($act\_id$) that reports value 1 for the row with the smallest value of $ms$. As the data matrix is sorted from the smallest $ms$ to the largest $ms$, a subsequent row ($t$) belongs to action 2 (i.e increases by 1) if $avg\_pos_x$ of the previous row ($t-1$) is on one side of the court and $avg\_pos_x$ in $t$ is on the other side. In doing that, we adopt a correction: $act\_id$ increases by 1 only if $avg\_pos_x$, from $t$ to $t+1$, passes over the transition area represented by the interval [+4,-4] meters from the half-court line (coloured area in Figure \ref{fig:5}).
\end{itemize}

After these two additional steps, the algorithm produces a reduced version of the data matrix \textbf{X} with action labelling and sorting. This generates a cleaned and ready-to-use data matrix that can be used for producing various kind of advanced statistics.

\section{Empirical Application}

In this section we test the algorithm on real basketball games. We also perform a calibration analysis for the choice of the parameters used in the step 1-C of the algorithm. After having applied it to the case studies (CS), we present some descriptive analysis to evaluate the performance of the algorithm in relation to specific elements that characterize the game, such as the number of actions and their duration. First of all, we present data and describe preliminary data processing that is needed before running the algorithm. 

\subsection{Data}

Data refer to three games played by Italian professional basketball teams, at the Italian Basketball Cup Final Eight. MYagonism  (\url{https://www.myagonism.com/}) was in charge to set up a system to capture these data during the games, trough accelerometer devices. Each player worn a microchip that, having been connected with machines built around the court, collected the player's position (in pixels of 1 $cm^2$ size) in the $x$-axis (court length), the $y$-axis (court width), and in the $z$-axis (how high the player jumps). Data have been detected with an average frequency of about 80 Hz. The initial data matrices contain information on players' positioning, velocity and acceleration during the full game length. Throughout the text, we will call the three games, Case Study 1 (CS1), Case Study 2 (CS2) and Case Study 3 (CS3)

\subsection{Data Processing}

\noindent Data have been provided to us in the form of several .csv files, each containing data of a single player in a single game. Each file is named with a univocal player's code, and a code-name concordance table has been provided.  Each file contains three variables: $label$, $ms$ and $value$. \textit{label} refers to the name of the captured information, that can be positioning, velocity or acceleration, in the $x$-, $y$- or $z$-axis.  \textit{ms} reports the millisecond while \textit{value} reports the information. Before the algorithm, preliminary data processing is needed. 

\begin{itemize}
	\item After having manually converted .csv files in .xls format, and having renamed files with the player's name, we generate several .txt files, each for a single player in a single game. These new files contains five columns: $label$, $ms$, $value$, $names$, $team$. $names$ reports the player's name. $team$ reports the team's name and the date of the game;
	\item we manipulate the files generated at the previous step in such a way the output file reports the variables positioning, acceleration and velocity, in $x$-, $y$- and $z$- axis, for each millisecond (variable $ms$) in which at least the data of one player has been captured. The data refer, in case a player has not been tracked in that millisecond, to the most recent available $ms$;
	\item $z$-axis and acceleration variables has been dropped, because not useful to the running of the algorithm. Then, $time$ (an identifier for sorted time instants) and $id$ (an identifier for the player) variables have been created. At this stage of the preliminary data processing, the data matrix is structured in order to draw motion charts, as described in \cite{metulini2017spatio}; 
	\item in the end, the data matrix has to be reshaped in such a way each player's variables  are reported in column. This data matrix is in the structure to be used as input for the algorithm. Positioning is expressed in $meters.centimeters$ from the half-court line, while velocity is expressed in meters per seconds ($m/s$).
\end{itemize}

\vline

A number of additional variables has been computed. These new variables are instrumental to validate the algorithm. In detail, we generate:

\begin{itemize}
	\item  The list $d_{1}$, ... , $d_{n^2-n}$ of $n^2-n$ variables reporting the distance (in $m$) between players' pairs, being the distance between player $i$ and player $j$ $d_{ij}=\sqrt{(pos_{p_i\_x} - pos_{p_j\_x})^2 + (pos_{p_i\_y} - pos_{p_j\_y})^2}$; the value of $d_{ij}$ is missing if at least one of the two players is on the bench in that specific $ms$;
	\item  the variable $d_{avg}$, which reports the average distance (in $m$) among all the $n^2-n$ distances ($d_{avg} = \frac{\sum_{i=1}^{n^2-n} d_i}{n^2-n}$);
	\item  the variable $con\_hull$, that reports the area (in $m^2$) occupied by the five players on the court, also called convex hull area;
	\item the variable $vel_{avg}$, which is the average velocity of the five players in the court ($vel_{avg} = \frac{\sum_{i=1}^{5} vel_i}{5}$).
\end{itemize}

We replicate this data manipulation procedure for each of the three CS.

\subsection{Parameters' Validation}

Recalling that the aim of the algorithm is to reduce the data matrix to a total of 40 minutes of game, we explore different combinations for the parameters $h_2$ and $h_3$ in relation to the resulting number of minutes. Minutes have been computed by comparing $ms$ in consecutive rows ($ms_t - ms_{t-1}$) in the non-reduced data matrix that enters the algorithm. As $ms$ is expressed in milliseconds, a proper conversion to minutes has made. 

\begin{figure}[htbp!]
	\centerline{ \includegraphics[width=0.45\linewidth]{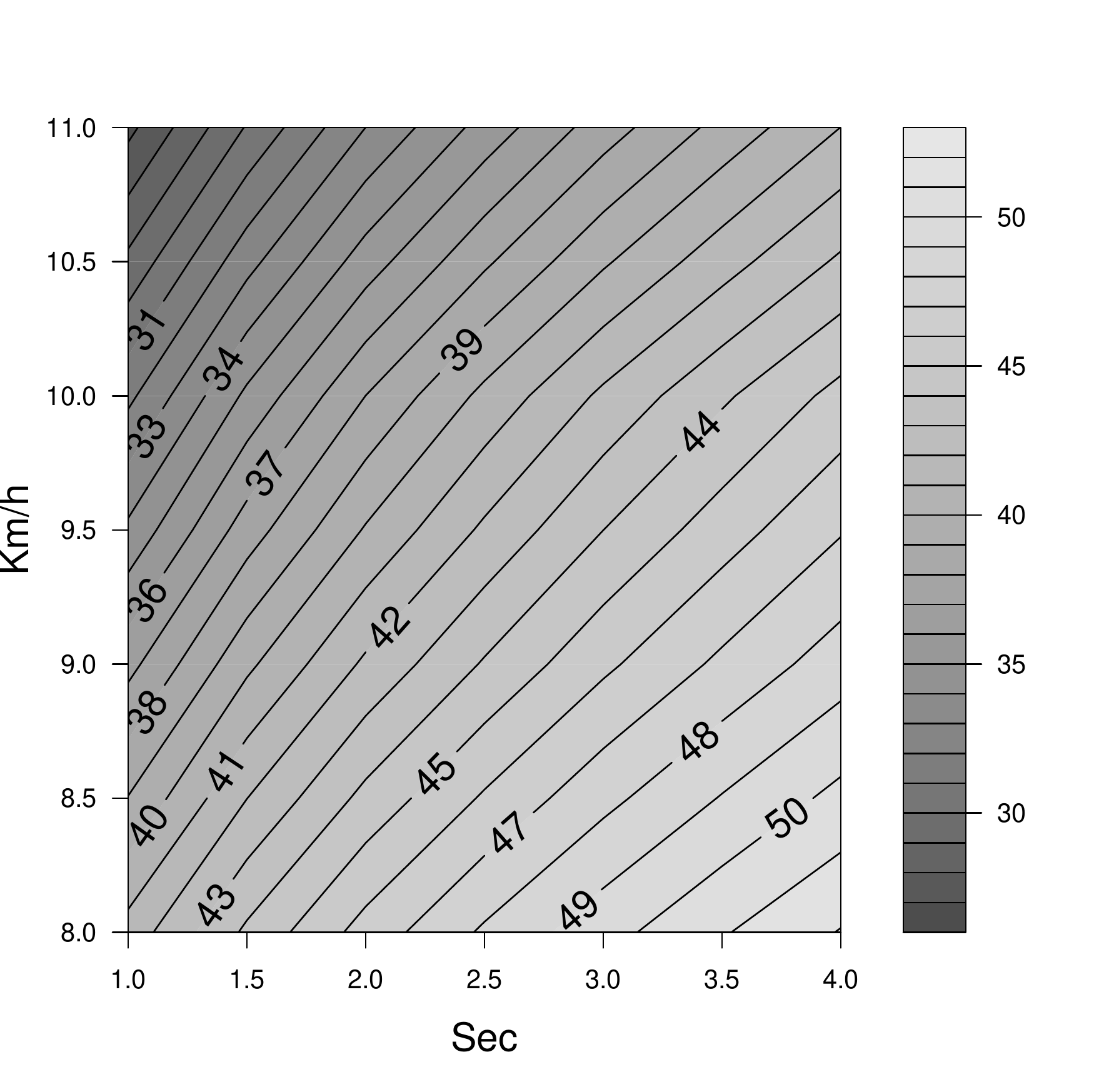}}
	\centerline{ \includegraphics[width=0.45\linewidth]{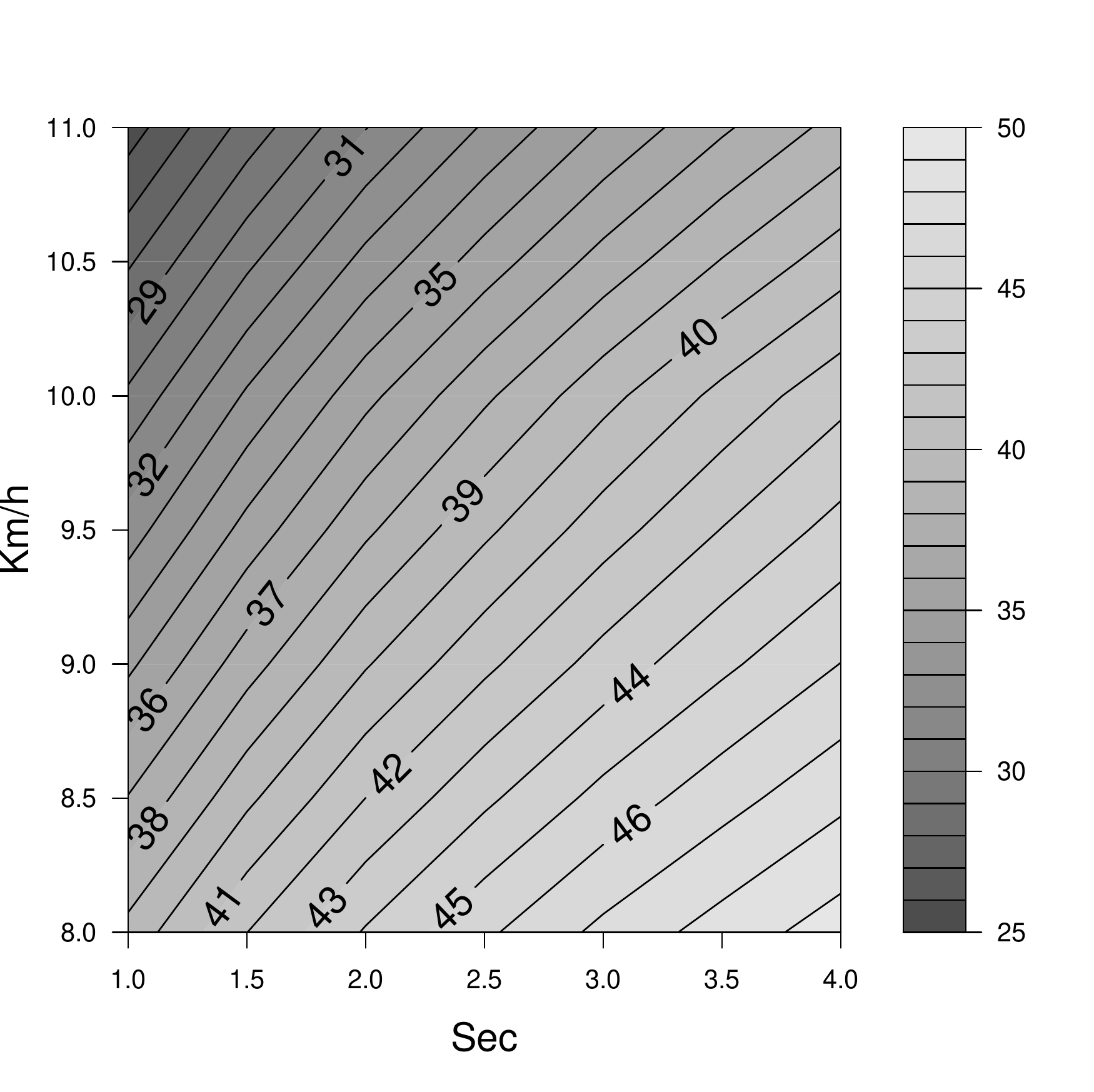}}
	\centerline{ \includegraphics[width=0.45\linewidth]{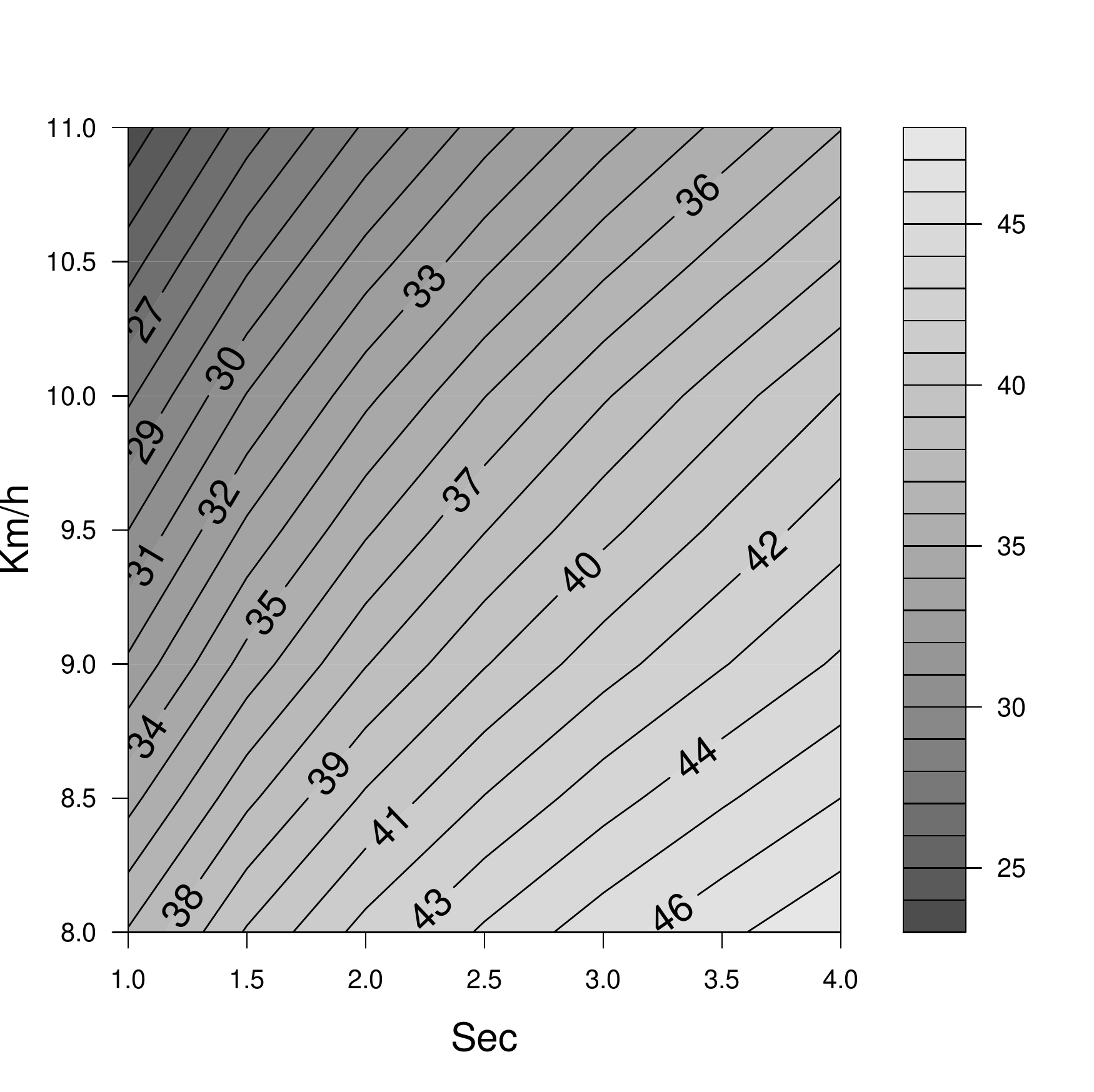}}
	\caption{\textit{Contour plot representing the effective minutes of game obtained by the algorithm, subject to different combination of parameters $h_2$ ($km/h$)  and $h_3$ (seconds) combination. From top to bottom: CS1, CS2 and CS3.}}
	\label{fig:6}       
\end{figure}

Moreover, since different games involve different team-mates and opponents, it is likely that the optimal combination for the parameters $h_2$ and $h_3$ slightly changes over CS1, CS2 and CS3. However, we do not want to find precise values, but a range of acceptable values.

Figure \ref{fig:6} displays contour plots for the three CS. The chart reports the contours levels for the length (in minutes) of the reduced data matrix, according to different combination of $h_2$ and $h_3$ parameters. The contours are evaluated on the range [8;11] according to the parameter $h_2$ and on the range [1;4] for $h_3$ parameter. 
We note that, for all the three CS, game length increases with the increase of $h_3$ parameter, while it decreases to an increase on the $h_2$ parameter.  
Interestingly, we find similar evidence over different CS, in terms of the choice of the parameters. Consistently over the three CS, the algorithm performs better when $h_2$ lies in the interval [9;10] and $h_3$ lies  in the interval [2;3]. 

\subsection{Results}

In this section we present the results of the algorithm for the three real case studies in terms of relevant game characteristics, such as the game length, the number of actions and the action's length distribution, as well as the average distances among team-mates and the area occupied by them. These characteristics could depend on coaches' tactics, players' strength and team-mates cohesion, but they may also depend on whether the action is an offensive or a defensive one, on the presence of a specific player or on a specific line-up in the court. 
Comparing game characteristics with those generally obtained in real games, serves as a test for the robustness of the algorithm. Based on the indication given by the parameter validation in the previous subsection, we choose $h_2 = 9$ $km/h$ and $h_3 = 2.5$ seconds for CS1, $h_2 = 9.4$ $km/h$ and $h_3 = 2.5$ seconds for CS2, $h_2 = 10$ $km/h$ and $h_3 = 2.5$ seconds for CS3.

\vline

\begin{figure}[h!]
\includegraphics[width=0.48\linewidth]{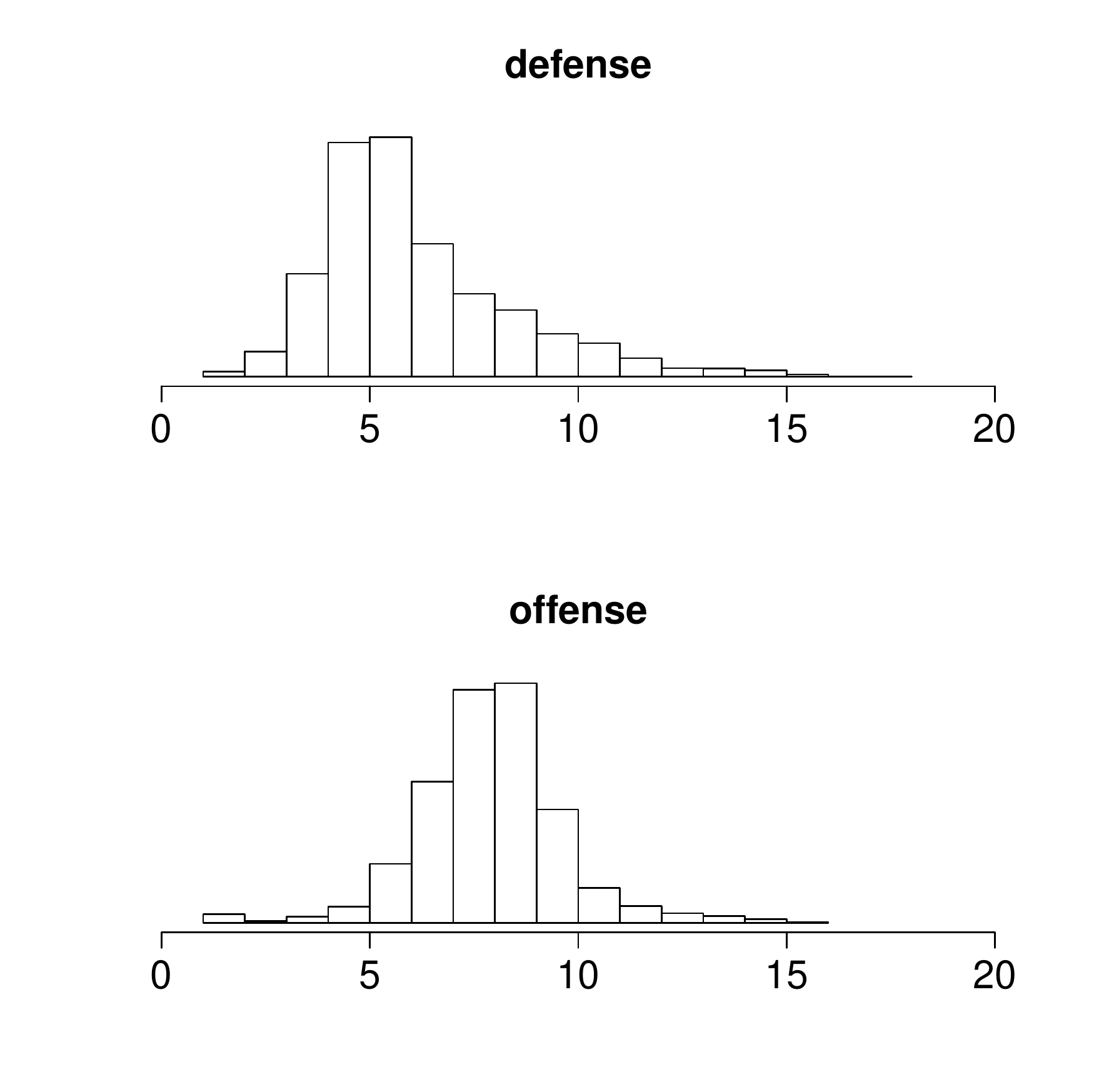}
		\includegraphics[width=0.48\linewidth]{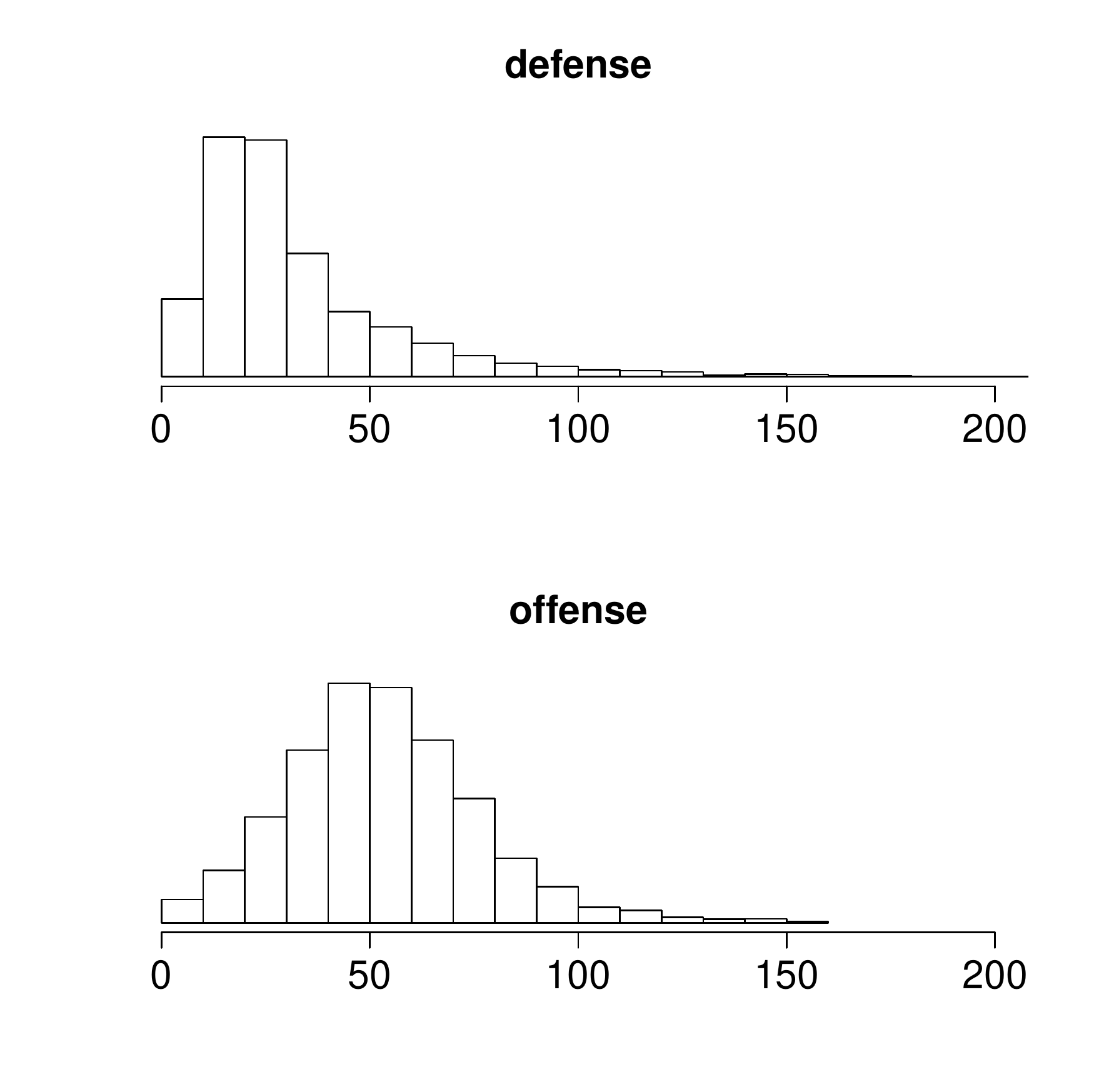}	
	\caption{Histogram distribution of average distances (in $m$, left) and average convex hull areas (in $m^2$, right) during offense and during defense, in CS1.}
	\label{fig:7}
\end{figure}

We start by characterizing the three games in terms of players' distances and surface area occupied by team-mates (convex hulls area). We report analyses separately for offensive and defensive instants, as \cite{metulini2017sensor} found that surface area significantly differs from offensive to defensive actions. Consistently with the results in the aforementioned work, average distance among players ($d_{avg}$), along the three case studies, is generally larger during offensive instants.  Left panel of Figure \ref{fig:7} reports as illustrative example the related distribution for CS1. In defense, the average distance among payers reports a mean of 6.17 $m$ and a median of 5.57 $m$. In offense, the average distance reports a mean of 7.96 $m$ and a median of 7.97 $m$. With reference to surface area, results give a similar information. Convex hulls area, along the three case studies, are generally larger during offense. Right panel of Figure \ref{fig:7} reports as illustrative example the related distribution for CS1. In defense, the convex hulls area reports a mean of 32.29 $m^2$ and a median of 24.97 $m^2$. In offense, the convex hulls area reports a mean of 53.46 $m^2$ and a median of 51.98 $m^2$. 

\begin{figure}[h!]
	\centering
	\includegraphics[width=0.8\linewidth]{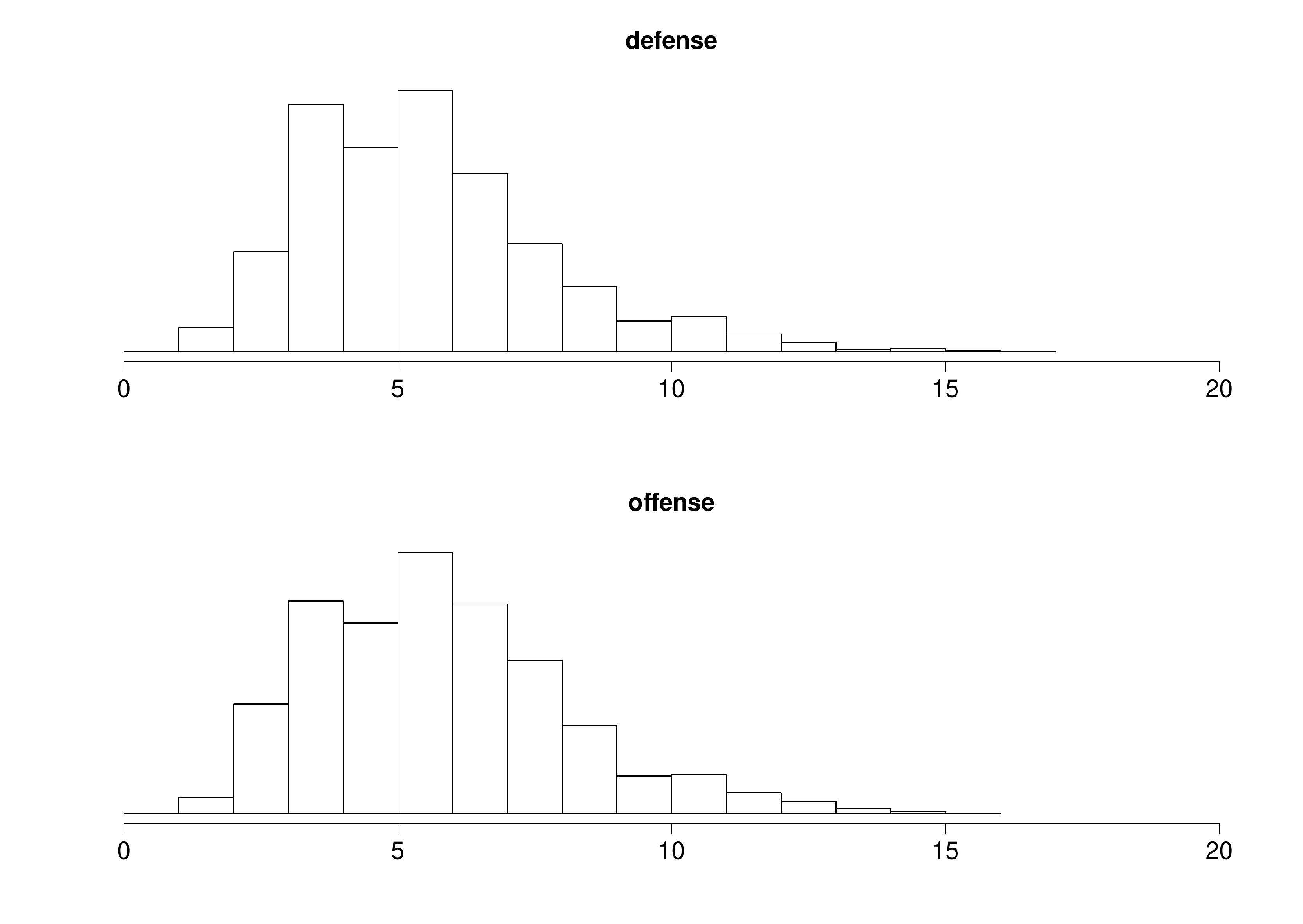}
	\caption{Histogram distribution of average velocity (in $km/h$) during offense and during defense, in CS1.}
	\label{fig:8}
\end{figure}

Alike, we computed the average velocity ($vel_{avg}$) of the five players on the court. Figure \ref{fig:8} reports the related distribution for CS1. In defense, the players' velocity reports a mean of 5.56 $km/h$ and a median of 5.10 $km/h$. In offense, the players' velocity reports a mean of 5.76 $km/h$ and a median of 5.44 $km/h$. Moreover, players' velocity is larger than 6.71 $km/h$ for the 25\% of the defensive moments, larger than 7.07 $km/h$ for the 25\% of the offensive moments.
 
\vline

We now introduce the distributions of the number of actions retrieved by the procedure. The algorithm, applied to real case studies, divides the game into a consistent number of actions. 
The algorithm splits the games in a number of 151 actions for CS1, 136 actions for CS2 and 132 for CS3. 

A robustness check consists on counting the number of actions that last for a reasonable time. Choosing an interval between 4 and 38 seconds\footnote{38 seconds, due to the new rule that gives 14 additional seconds after a foul, and also considering a double ball possession after an offensive rebound.}, in CS1 147 out of 151 actions are included in this interval (97.4\%). This number is 133, out of 136 (97.8\%), for CS2 and 128, out of 132 (97.0\%), for CS3. All in all, it emerges that most of the actions lie in a reasonable interval of duration, consistently along all the three games analysed.

\begin{figure}[h!]
	\centering
	\includegraphics[width=0.8\linewidth]{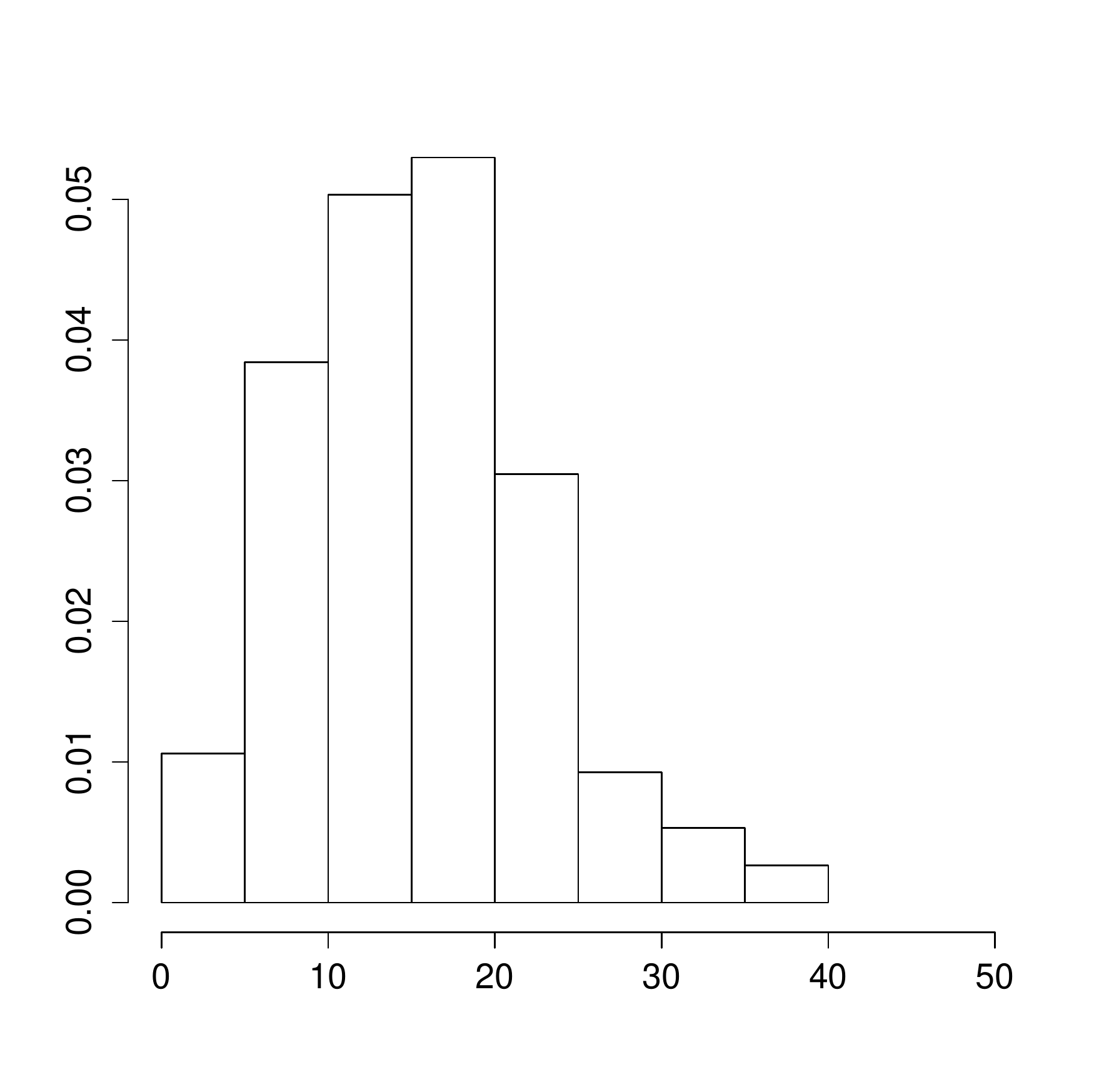}
	\caption{Histogram distribution representing the action duration (in seconds), in CS1.}
	\label{fig:9}
\end{figure}

The histogram in Figure \ref{fig:9} reports the distribution of the actions' duration in CS1. This histogram shows that actions last for a reasonable interval of time. The average duration stands to 16.00 seconds, the median stands to 15.66. Furthermore, the 23.18\% of the actions last for less then 10 seconds, the 50.33\% of them last for 10 to 20 seconds and the 26.49\% last for more than 20 seconds.   

\vline

Does the algorithm produce good results in terms of retrieving the correct number of actions in a game and the correct action duration? Until now, we have described the results obtained by applying the algorithm to three real case studies, and by evaluating them with respect to the \textit{rule of common sense}. 

\begin{figure}[htbp!]
	\includegraphics[width=0.5\linewidth]{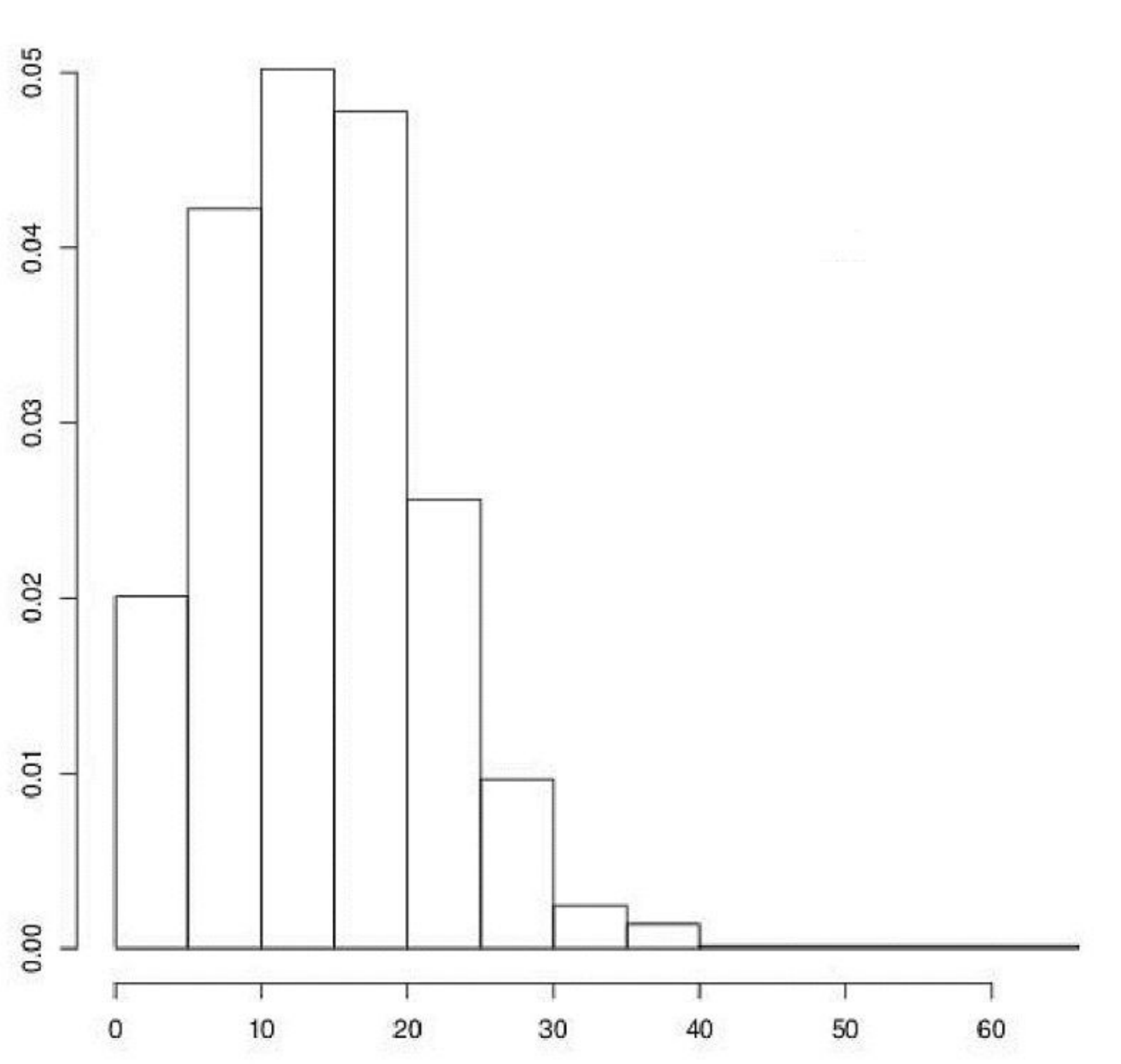}
	\includegraphics[width=0.5\linewidth]{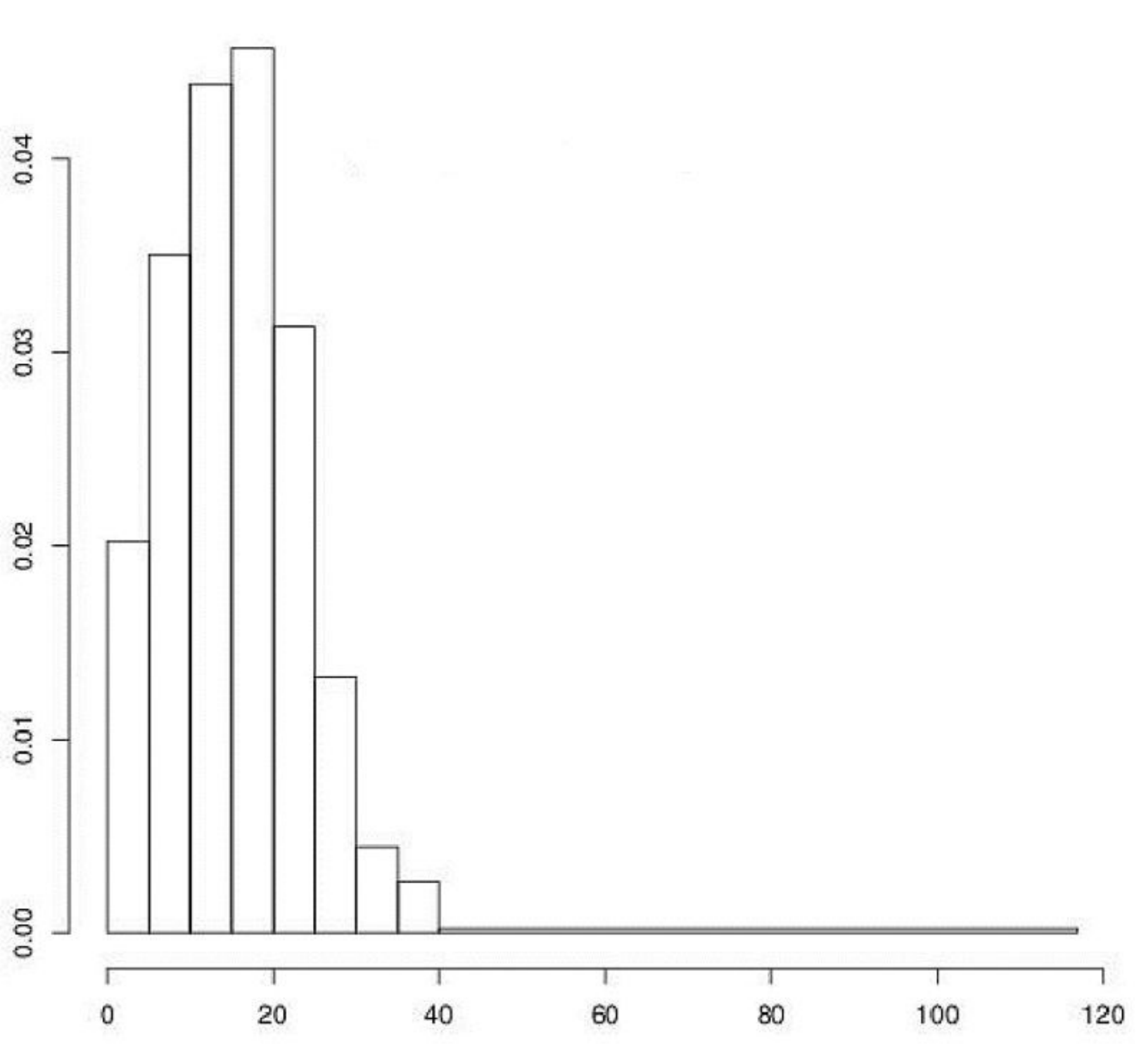}
	\caption{Action duration for team USA during Olympic Games (Left) and for team Leonessa Brescia (Right). Data reports information on aggregated games.}
	\label{fig:10}
\end{figure}

Now, in order to further validate the procedure, we compare values from applying the algorithm to real CS with true information on real games. These information has been retrieved by looking to the web-scraped play-by-play of several games from both 2016 FIBA Olympics and 2015-2016 Italian professional \textit{A2} tournament. In relation to a bunch of matches, we report the distribution of the number of actions per game and the  distribution of the actions' duration. More specifically, we use, as an example, one national team (USA) during the Olympic games and one Italian team (Leonessa Brescia) from \textit{A2}. Histograms reported in Figure \ref{fig:10} refer to aggregated information from several games (i.e. the national team played 8 times during the Olympic games, \textit{A2} tournament lasts for 30 games.). Other teams of the same leagues displays similar results.

Left panel of Figure \ref{fig:10} displays the distribution of the actions' duration (in seconds) for team USA, while right panel reports the same distribution for team Leonessa Brescia. The two distributions are similar. The median duration for team Leonessa Brescia stands to 16. The median duration for team USA stands to 14. Moreover, the distributions present only few actions with a duration smaller than 10 seconds (31.60\% for team USA, 27.50\% for team Leonessa Brescia) or larger then 20 seconds (19.40\% for team USA, 25.20\% for team Leonessa Brescia), while most of the actions last for 10 to 20 seconds (49.00\% for team USA, 47.30\% for team Leonessa Brescia). These values are similar to those from applying the algorithm to CS1.

\begin{figure}[htbp!]
	\includegraphics[width=0.5\linewidth]{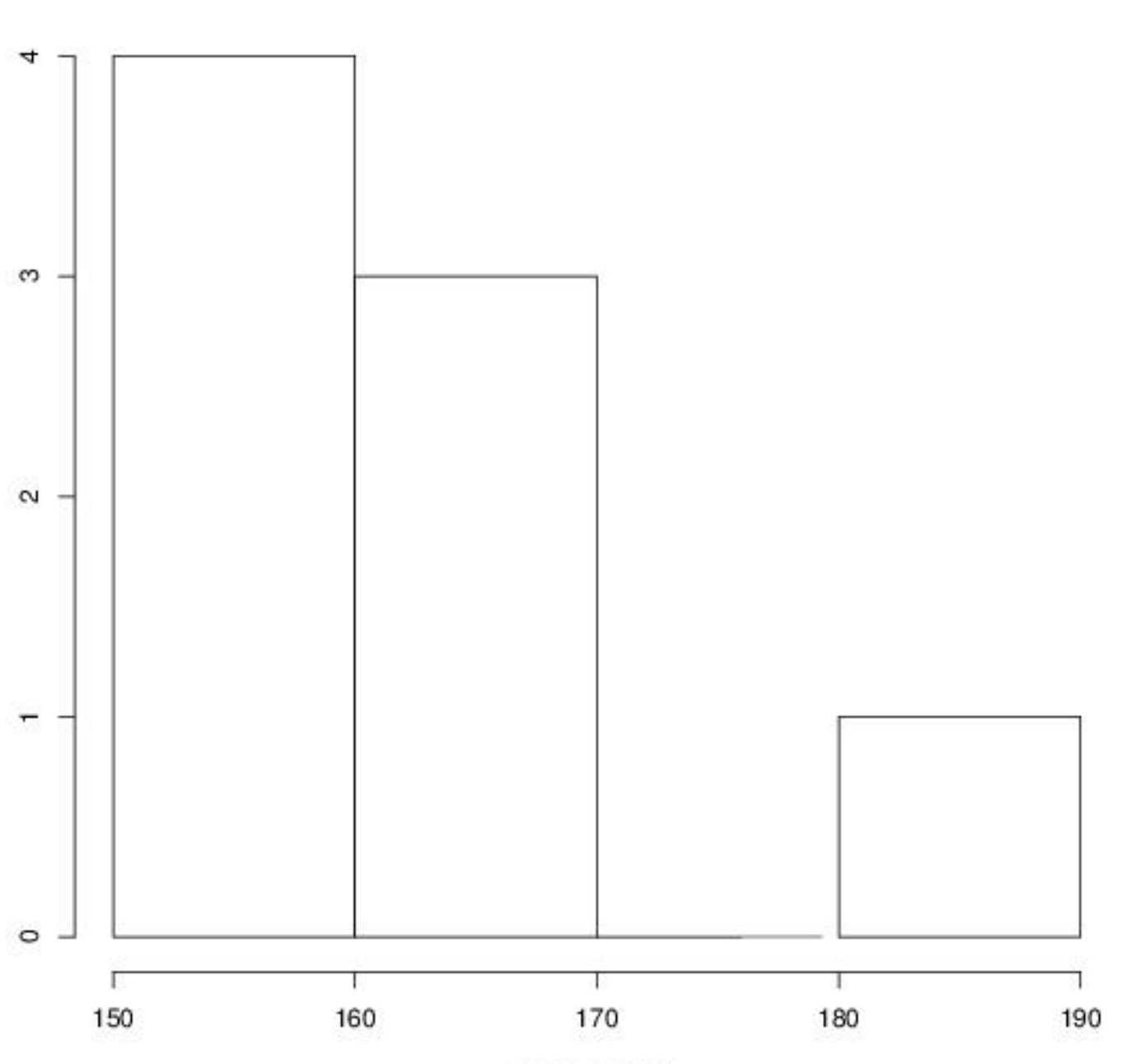}
	\includegraphics[width=0.5\linewidth]{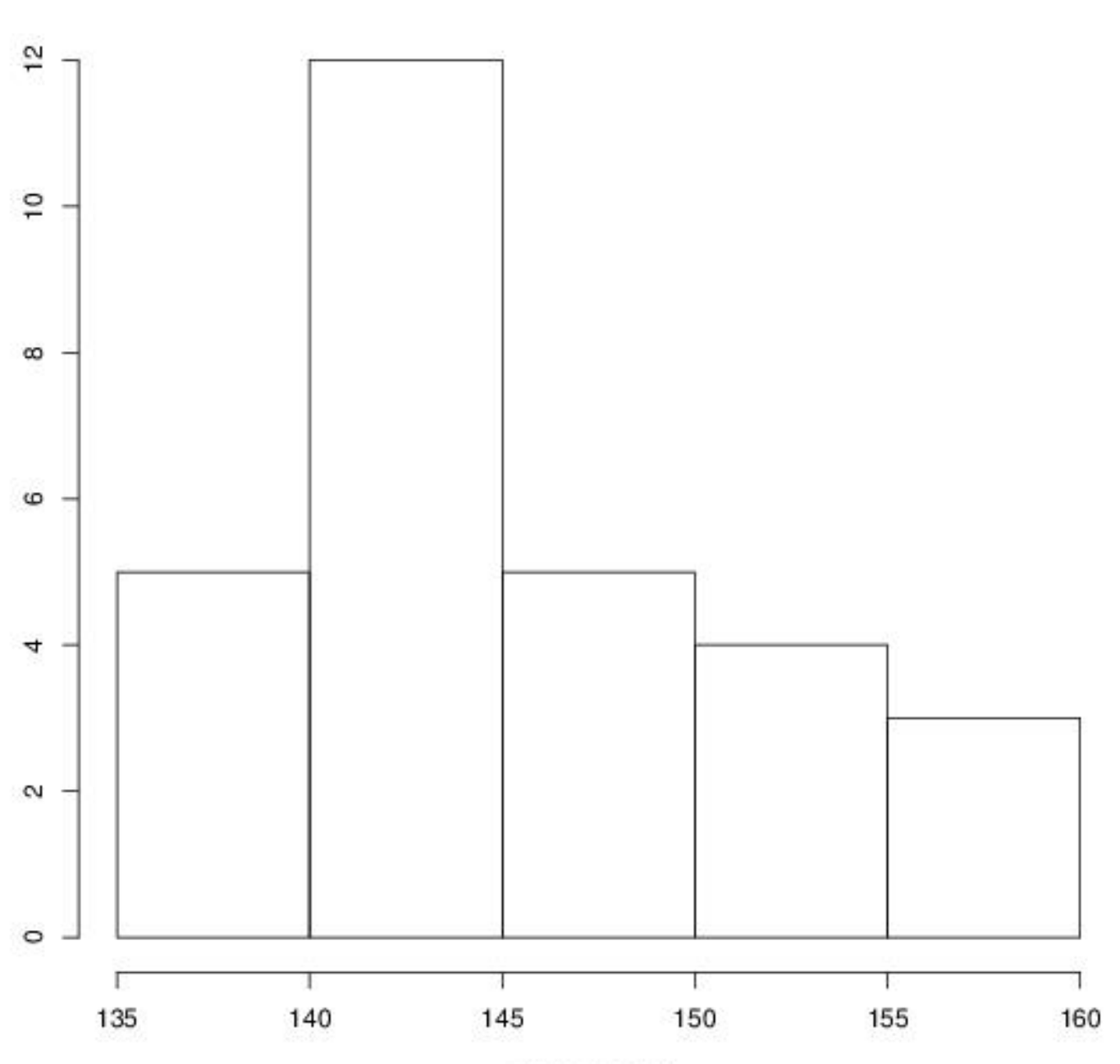}
	\caption{Number of actions for team USA during Olympic Games (Left) and for team Leonessa Brescia (Right). Data reports information on aggregated games.}
	\label{fig:11}
\end{figure}

Left panel of Figure \ref{fig:11} displays the distribution of the number of actions in a single game, for team USA. Right panel reports the same distribution for team Leonessa Brescia. For the latter, the game lasted 140 to 150 actions 17 out of 29 times (58.62\%). For team USA, the game lasted 150 to 170 actions 7 out of 8 times (87.5\%).

All in all, these values are consistent with those obtained by applying the algorithm to the real CS and demonstrate the robustness of our procedure. 

\section{Discussion and Conclusions}

In the era of the Information Technology and Big Data Analytics, team sports' managers benefit from the availability of advanced statistics. However, statistics are just a tip in the iceberg, while there is an hard work behind, which concerns the steps of tracking, collecting, storing and processing the data. This paper concerned with basketball data processing, and aimed to suggest an ad-hoc procedure to automatically filter a data matrix containing players' movement information to the moments in which the game is active, and by dividing the game into sorted and labelled actions. In this regard, we placed this work within the area of the Human Activity Recognition, as we used players' actions to recognize a specific game state (i.e. inactive game moments).  The algorithm has been tested on three different real games, and a series of robustness checks has been done, including a validation for the parameters to be used in the algorithm. Results of the validation suggests a stability of the two parameters along different games. 

Practitioners which are in possession of a data matrix as the one described in this paper can replicate this procedure to analyse basketball matches. 

The novelty of this procedure is that, unlike existing works, for example \cite{wu2017modeling}, it works when the ball's trajectory is unavailable. However, further research is to be planned in order to validate the algorithm also with respect to a visual analysis of the same match. 

\section*{Acknowledgments}
Research carried out in collaboration with the Big\&Open Data Innovation Laboratory (BODaI-Lab), University of Brescia (project nr. 03-2016, title Big Data Analytics in Sports, www.bodai.unibs.it/BDSports/), granted by Fondazione Cariplo and Regione Lombardia. The author thanks Paola Zuccolotto, Marica Manisera (University of Brescia) and Tullio Facchinetti (University of Pavia) for valuable suggestions.

%
%

\begin{thebibliography}{99.}%
%
%
\bibitem{Araujo2016team} Ara\`ujo, D., Davids, K. Team synergies in sport: theory and measures. {\it Frontiers in psychology}, \textbf{7} (2016), 1449.

\bibitem{carpita2013} Carpita, M., Sandri, M., Simonetto, A., Zuccolotto, P.: Football mining with R. Data Mining Applications with R (2013).

\bibitem{carpita2015} Carpita, M., Sandri, M., Simonetto, A., Zuccolotto, P.: Discovering the Drivers of Football Match Outcomes with Data Mining. Quality Technology \& Quantitative Management 12.4 (2015): 561-577.

\bibitem{gudmundsson2017spatio} Gudmundsson, J., Horton, M.: Spatio-temporal analysis of team sports. \textit{ACM Computing Surveys (CSUR)}, \textbf{50}(2) (2017), 22.

\bibitem{huang2012calculate} Huang, Y. C., Chen, T. L., Chiu, B. C., Yi, C. W., Lin, C. W., Yeh, Y. J., Kuo, L. C.: Calculate golf swing trajectories from imu sensing data. \textit{In Parallel Processing Workshops (ICPPW), 2012 41st International Conference on} (2012) (pp. 505-513). IEEE.

\bibitem{jiang2004new} Jiang, S., Ye, Q., Gao, W., Huang, T.: A new method to segment playfield and its applications in match analysis in sports video. \textit{In Proceedings of the 12th annual ACM international conference on Multimedia} (2004) (pp. 292-295). ACM.

\bibitem{jordan2009optimizing} Jordan, J. D., Melouk, S. H., Perry, M. B.: Optimizing football game play calling. \textit{Journal of Quantitative Analysis in Sports}, (2009) \textbf{5}(2).

\bibitem{metulini2017spatio} Metulini, R.: Spatio-temporal movements in team sports: a visualization approach using motion charts. \textit{Electronic Journal of Applied Statistical Analysis}, (2017) \textbf{10}(3).

\bibitem{metulini2017sensor} Metulini, R., Manisera, M., Zuccolotto, P.: Sensor Analytics in Basketball. \textit{Proceedings of the 6th International Conference on Mathematics in Sports.} (2017).

\bibitem{passos2011networks} Passos, P., Davids, K., Araujo, D., Paz, N., Minguens, J., Mendes, J.: Networks as a novel tool for studying team ball sports as complex social systems. Journal of Science and Medicine in Sport 14.2 (2011): 170-176. 

\bibitem{stein2017sense} Stein, M., Janetzko, H., Seebacher, D., Jager, A., Nagel, M., Holsch, J., Grossniklaus, M.: How to make sense of team sport data: From acquisition to data modeling and research aspects. \textit{Data}, (2017) \textbf{2}(1), 2.

\bibitem{tinder2006flight} Tinder, R. F.: Relativistic flight mechanics and space travel. \textit{Synthesis lectures on engineering}, (2006) \textbf{1}(1), 1-140.

\bibitem{trabelsi2013approach} Trabelsi, D., Mohammed, S., Chamroukhi, F., Oukhellou, L., Amirat, Y.: An unsupervised approach for automatic activity recognition based on hidden Markov model regression. \textit{IEEE Transactions on automation science and engineering}, (2013) \textbf{10}(3), 829-835.

\bibitem{travassos2013performance} Travassos, B., Davids, K., Araujo, D., Esteves, P. T.: Performance analysis in team sports: Advances from an Ecological Dynamics approach. International Journal of Performance Analysis in Sport 13.1 (2013): 83-95.

\bibitem{wasserman1994social} Wasserman, S., Katherine F.: Social network analysis: Methods and applications, Vol. 8. Cambridge university press (1994). 

\bibitem{wu2017modeling} Wu, S., Bornn, L.: Modeling offensive player movement in professional basketball.\textit{ The American Statistician.} (2017).


%
\end{thebibliography}
%

\end{document}